\newif\ifonecolumnlayout

\onecolumnlayoutfalse 

\ifonecolumnlayout
  \PassOptionsToClass{onecolumn,notitlepage,11pt}{revtex4-2}
\else
  \PassOptionsToClass{twocolumn,10pt}{revtex4-2}
\fi
\PassOptionsToClass{showkeys,nofootinbib,amsmath,amssymb,aps,pra,floatfix}{revtex4-2}
\documentclass{revtex4-2}

\ifonecolumnlayout
  \usepackage[a4paper,margin=2.5cm]{geometry}
\fi

\usepackage[T1]{fontenc}
\usepackage{newtxtext,newtxmath}
\usepackage{titlesec}
\usepackage{soul}
\usepackage{xcolor}

\usepackage{empheq}
\usepackage[english]{babel}
\usepackage{url}
\usepackage{silence}
\makeatletter
\let\label\ltx@label
\makeatother
\usepackage[colorlinks=true,linkcolor=blue,citecolor=red,urlcolor=blue,filecolor=blue,linktocpage=true]{hyperref}
\hypersetup{allcolors=blue,citecolor=red}
\usepackage{cleveref}
\usepackage[utf8]{inputenc}
\usepackage[toc,page]{appendix}
\usepackage[autostyle]{csquotes}
\usepackage{slashed}
\usepackage{graphicx}
\usepackage[caption=false]{subfig}
\usepackage{dcolumn}
\usepackage{longtable}
\usepackage{array}
\usepackage{bm} 
\usepackage{amsmath}
\makeatletter
\let\over\@@over
\makeatother
\usepackage{nccmath}
\usepackage{cancel}
\usepackage{mathtools}
\usepackage{amsfonts}
\usepackage{upgreek}
\usepackage{paralist} 
\DeclareMathAlphabet{\mathpzc}{OT1}{pzc}{m}{it} 
\usepackage{tensor}
\makeatletter
\renewcommand{\@keys@name}{\textsc{Keywords: }}
\makeatother

\titleformat{\section}{\normalfont\large\bfseries\sffamily}{\thesection}{1em}{}
\titleformat{\subsection}{\normalfont\normalsize\bfseries\sffamily}{\thesubsection}{1em}{}
\titleformat{\subsubsection}{\normalfont\normalsize\bfseries\sffamily}{\thesubsubsection}{1em}{}
\titleformat{\paragraph}[runin]{\normalfont\normalsize\bfseries\sffamily}{\theparagraph}{1em}{}[.]
\titlespacing*{\section}{0pt}{2.8ex plus 0.8ex minus .2ex}{1.1ex plus .2ex}
\titlespacing*{\subsection}{0pt}{2.4ex plus 0.8ex minus .2ex}{1.1ex plus .2ex}
\titlespacing*{\subsubsection}{0pt}{2.4ex plus 0.8ex minus .2ex}{1.1ex plus .2ex}
\titlespacing*{\paragraph}{0pt}{1.5ex plus 0.5ex minus .2ex}{1em}
\renewcommand{\thesection}{\arabic{section}}
\renewcommand{\thesubsection}{\thesection.\arabic{subsection}}
\renewcommand{\thesubsubsection}{\thesubsection.\arabic{subsubsection}}

\makeatletter
\def\Dated@name{}

\def\frontmatter@above@affiliation{\par\vspace{10pt}}
\def\frontmatter@above@affilgroup{\par\vspace{10pt}}
\makeatother

\makeatletter
\renewcommand{\@dotsep}{10000}
\renewcommand\l@section[2]{\vskip 0.45em\@dottedtocline{1}{0em}{1.8em}{#1}{{\fontseries{b}\selectfont #2}}}
\renewcommand\l@subsection[2]{\@dottedtocline{2}{1.8em}{2.2em}{#1}{#2}}
\renewcommand\l@subsubsection[2]{\@dottedtocline{3}{4.0em}{2.8em}{#1}{#2}}
\makeatother

\makeatletter
\def\ps@jhepfooter{%
  \let\@oddhead\@empty
  \let\@evenhead\@empty
  \def\@oddfoot{\hfil--\ \thepage\ --\hfil}%
  \def\@evenfoot{\hfil--\ \thepage\ --\hfil}%
}
\makeatother

\begin{document}
\hbadness=10000
\sloppy

\addtolength{\footskip}{6pt}
\pagestyle{jhepfooter}

\ifonecolumnlayout\preprint{}\else\preprint{APS/123-QED}\fi

\title{{\sffamily\bfseries\Large New-born strings are tensionless}}

\author{Sudip Karan}
\author{and Bibhas Ranjan Majhi}
\affiliation{ Department of Physics, Indian Institute of Technology Guwahati, Guwahati 781039, Assam, India}
\affiliation{\vspace{7pt}\textit{E-mail:} {\normalfont\href{mailto:sudip.karaan@gmail.com}{sudip.karaan@gmail.com}}, {\normalfont\href{mailto:bibhas.majhi@iitg.ac.in}{bibhas.majhi@iitg.ac.in}}}

\date{}

\begin{abstract}
\noindent We report a physical origin of tensionless strings, obtained by formulating string dynamics in finite-lifetime settings for the first time. Departing from the conventional paradigm of eternal strings, we construct an inertial worldsheet confined to a finite region of two-dimensional Minkowski spacetime, known as a causal diamond. This reveals a striking result: tensionless strings arise only at the moment of their birth. The ultra-shrinking limit of the diamond worldsheet realizes this birth configuration, thereby uncovering a new tensionless string phase characterized by a global, ultra-local Carrollian structure.
\end{abstract}

\maketitle
\ifonecolumnlayout
  \thispagestyle{empty}\setcounter{page}{0}\newpage
\else
  \setcounter{page}{1}\thispagestyle{jhepfooter}
\fi

\begingroup
\hypersetup{linkcolor=blue}
\setlength{\parskip}{0pt}
\endgroup
\hypersetup{linkcolor=blue}

	
\section{Introduction}\label{intro}

\begin{figure}[ht]
\centering
\includegraphics[width=1.0\columnwidth]{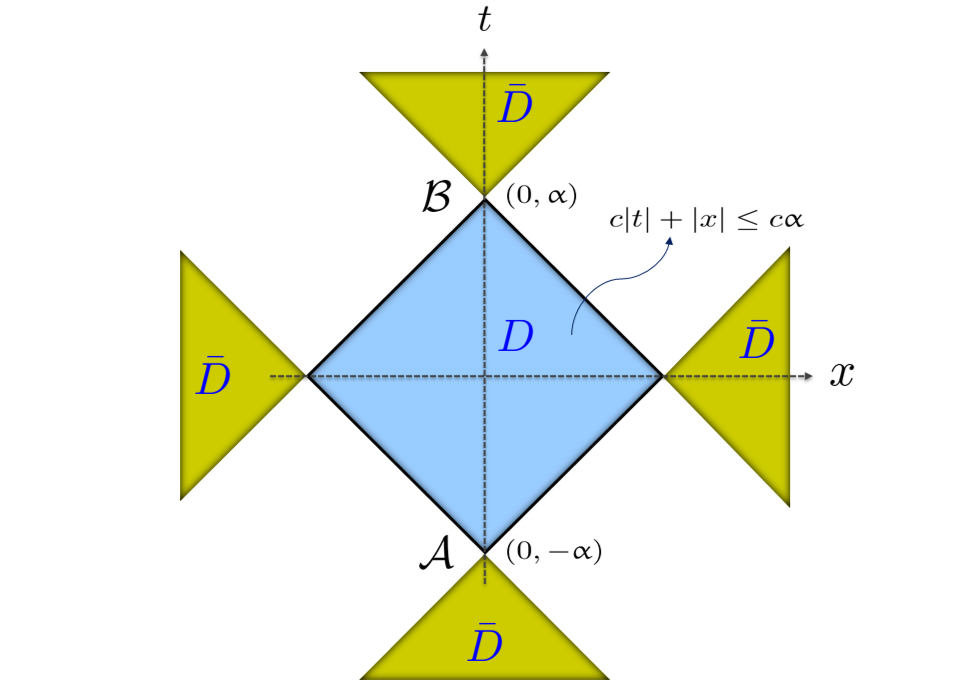}
\caption{Causal diamond in (1+1)-dimensional Minkowski spacetime, bounded by the birth event at $\mathcal{A}(0,-\upalpha)$ and the death event at $\mathcal{B}(0,\upalpha)$. The finite lifetime, set by the diamond size $\mathcal{AB}=2\upalpha$, confines the observer to the diamond interior region $D$ and restricts access to the exterior regions $\bar{D}$.}\label{fig:cd}
\end{figure}

The study of physics in causally restricted regions of spacetime has been a remarkably fruitful avenue in relativistic quantum field theory. Observers with limited causal access perceive horizons and experience thermality, with the Minkowski vacuum appearing as a mixed state due to tracing over inaccessible degrees of freedom \cite{Wald:1995yp,Birrell:1982ix}. This is most prominently realized in the Unruh effect, where accelerated observers restricted to a Rindler wedge detect a thermal spectrum \cite{Unruh:1976db}. In modern parlance, such features are not tied to acceleration alone, but arise whenever the observer is confined to any causally restricted domain, including the future and past regions of Minkowski spacetime \cite{Olson:2010jy}. A particularly sharp realization occurs for observers with finite lifetime, for whom the accessible region is a causal diamond rather than an unbounded wedge (see \cref{fig:cd}), and even inertial observers perceive thermality governed entirely by causal structure \cite{Martinetti:2002sz,Martinetti:2008ja,Chakraborty:2022qdr,DeLorenzo:2017tgx,Jacobson:2018ahi}. In such cases, the Unruh response originates from the restriction of access to quantum degrees of freedom within the diamond, leading to a nontrivial relation between global and diamond-local modes encoded through Bogoliubov transformations and an associated entangled vacuum structure \cite{Casini:2011kv,Su:2015oys,Ida:2012um,Camblong:2024bsy,Foo:2020rsy}.

Our discussions in this paper are rather distinct. We seek to understand how such physics and the anatomy of finite causal restrictions on inertial observers manifest on the worldsheet of a closed string, and in particular, how they influence the structure of string dynamics. String theory is typically formulated with an implicit assumption of an eternal worldsheet, where the string has unrestricted access to the full spacetime \cite{Polchinski:1998rq}. Here, we depart from this paradigm and instead lay the foundation for a finite-lifetime worldsheet, obtained by restricting the string propagation to a background causal diamond geometry, thereby promoting causal accessibility to a dynamical ingredient in the worldsheet description. This perspective draws some inspiration from developments in non-inertial and horizon-induced yet eternal worldsheet constructions \cite{Bagchi:2020ats,Bagchi:2021ban,Karan:2024jgn}.

The motivation for doing so is closely tied to the tensionless limit of string theory \cite{Schild:1976vq,Isberg:1993av}. In this regime, the worldsheet becomes null and the dynamics reorganize into a qualitatively distinct phase governed by 3d Bondi–Metzner–Sachs ($\mathrm{BMS}_3$) symmetry algebra, replacing the usual Virasoro structure \cite{Bagchi:2013bga,Bagchi:2015nca}. Understanding the physical origin of this regime is therefore crucial, as it probes the ultra-high-energy regime of string theory \cite{Gross:1987ar,Gross:1987kza,Gross:1988ue} and encodes enhanced symmetry structures underlying its dynamics \cite{Bagchi:2026wcu}. Conventionally, this limit is introduced either through explicit scaling of the string tension or through geometric settings where null directions arise, such as near-horizon descriptions, where the emergence of tensionless dynamics is intimately tied to horizon physics (e.g. \cite{Bagchi:2020ats,Bagchi:2021ban,Bagchi:2023cfp}). More broadly, these structures are deeply connected to the underlying Carrollian geometry and its associated symmetry algebras \cite{Levy-Leblond:1965dsc,SenGupta:1966qer,Bacry:1968zf,Henneaux:1979vn} (see also \cite{Duval:2014uva,Duval:2014uoa,deBoer:2021jej,Samanta:2025lcq}). However, such realizations typically rely on asymptotic or horizon-based limits and therefore lie beyond directly observer-accessible regimes \cite{Bagchi:2026wcu}.

The causal diamond worldsheet offers a different perspective. The presence of a finite lifetime introduces an intrinsic scale on the worldsheet and naturally leads to two distinct regimes. One corresponds to the boundary of causal accessibility, marking the approach to the “death” of the worldsheet and paralleling standard horizon-based analyses. The other arises in the opposite extreme, where the worldsheet collapses to zero lifetime, leading to a degenerate configuration that is interpreted as the “birth” of the worldsheet. This setup allows us to examine whether the emergence of null dynamics—and their associated Carrollian structure—is confined to horizon limits, or whether it can arise more generally from the geometric structure and physical phase of the worldsheet itself.

In this work, we develop a framework to analyze string dynamics in such finite-lifetime settings. Building on analogies with Unruh-type constructions, we formulate the mode expansion on the causal diamond worldsheet and study the relation between global and localized modes, together with the structure of the corresponding vacuum. This provides a natural setting to analyze how causal restriction induced by finite lifetime reorganizes worldsheet dynamics and to probe the emergence of tensionless behavior beyond conventional horizon-based scenarios. In particular, as the diamond worldsheet approaches its birth state, the theory exhibits a reorganization toward a tensionless regime characterized by a globally degenerate and null worldsheet structure. While this behavior closely mirrors the tensionless dynamics traditionally associated with horizon limits, the present construction instead reveals a fundamentally different origin, tied to the collapse of the causal domain itself. In this sense, it follows that the tensionless string phase need not be exclusively associated with worldsheet horizon physics, but may instead emerge intrinsically at the birth of the worldsheet, indicating a hitherto unexplored global, ultra-local origin of Carrollian structure.

\section{Causal diamond on the worldsheet} \label{sec1}

This section develops the technical framework for a string worldsheet restricted to a causal diamond region of Minkowski spacetime. We construct the corresponding quantum mode expansion for a finite-lifetime worldsheet and determine the modifications induced by the causal diamond structure. A central element of the construction is a Unruh-type framework \cite{Unruh:1976db}, which enables the derivation of Bogoliubov transformations relating the diamond worldsheet modes and oscillation operators to those of the eternal Minkowski worldsheet. The analysis follows analogous constructions in causal diamond quantum field theory \cite{Su:2015oys,Ida:2012um,Camblong:2024bsy}, now extended to the string-theoretic setting where the interplay of worldsheet modes, oscillators, and vacuum structure becomes intrinsically richer.

\subsection{The setup}\label{setup}

The worldsheet dynamics of a string propagating in a Minkowski target space with metric $\eta_{\mu\nu}$ are governed by the Polyakov action \cite{Polchinski:1998rq}
\begin{align}\label{S1}
\tensor*{\mathcal{S}}{_{\rm P}} = -{\frac{\mathcal{T}}{2}}\int d\tau d\sigma \sqrt{-\gamma}\, \gamma^{ab}\partial_aX^{\mu}\partial_bX^{\nu}\eta_{\mu\nu}.
\end{align} 
In the conventional formulation, the string is assumed to have infinite lifetime and constant tension $\mathcal{T} = {1 \over 2\pi\alpha^\prime}$. Fixing the worldsheet metric to the conformal gauge $\gamma_{ab} = \eta_{ab}$ reduces the equations of motion to the two-dimensional wave equation in the worldsheet coordinates $\sigma^a = (\tau,\sigma)$, with $a,b = 0,1$. Accordingly, the most general solution can be recast into a form that precisely mirrors the familiar quantum mode expansion of a massless Klein–Gordon scalar field \cite{Karan:2024jgn}:
 \begin{align}\label{S2.1}
	X^\mu (\tau, \sigma) =& x^\mu + \alpha^\prime p^\mu\tau \nonumber \\
  &+ \sqrt{2\pi\alpha^\prime}\sum_{n > 0} \left[ \tilde{\alpha}_n^\mu \tilde{f}_n + \alpha_n^\mu f_n + \tilde{\alpha}_{-n}^\mu \tilde{f}_n^* + \alpha_{-n}^\mu f_n^* \right].
\end{align} 
Here, $x^\mu$ and $p^\mu$ respectively represent the center-of-mass position and momentum of the string, which together constitute the zero-mode sector ($n=0$). For a closed string with periodicity $\sigma \sim \sigma + \ell$, the right- and left-moving mode functions $\tilde{f}_n$ and $f_n$, together with their Hermitian conjugates $\tilde{f}_n^*$ and $f_n^*$, take the form \cite{Karan:2024jgn}:
\begin{align}\label{S2.2}
\tilde{f}_n = {i  \over \sqrt{4\pi}n}e^{-i\omega_n (\tau - \sigma/c)}, \quad f_n = {i  \over \sqrt{4\pi}n}e^{-i\omega_n (\tau + \sigma/c)},
\end{align}
where $\omega_n = {2\pi cn \over \ell}$ are the mode frequencies. The expansion \eqref{S2.1} therefore involves an infinite set of oscillators: ${\tilde{\alpha}_n^\mu,  \alpha_n^\mu}$ act as annihilation operators, while ${\tilde{\alpha}_{-n}^\mu, \alpha_{-n}^\mu}$ serve as creation operators.\footnote{The quantum mode expansion in \cref{S2.1,S2.2} and subsequent analysis forbid the replacement $n \to -n$ in worldsheet mode functions or oscillators. This enforces consistency with the structure of quantum field theoretic mode expansions \cite{Birrell:1982ix} and stands in marked contrast to the standard practice in the string theory literature \cite{Polchinski:1998rq}.} These operators satisfy the canonical commutation relations and define the Minkowski worldsheet vacuum $\lvert 0_{\rm M}\rangle$,    
\begin{align}\label{S3}
\begin{gathered}
	\left[\tilde{\alpha}_n^\mu, \tilde{\alpha}_m^\nu\right] =  \left[{\alpha}_n^\mu, {\alpha}_m^\nu\right] = n\eta^{\mu\nu}\delta_{n+m,0}, \\[2pt] 
  \tilde{\alpha}_n^\mu\lvert0_{\rm M}\rangle = {\alpha}_n^\mu\lvert0_{\rm M}\rangle = 0.
  \end{gathered}
\end{align}
We now consider a worldsheet whose target space is restricted to a causal diamond region. This amounts to replacing the Minkowski worldsheet \eqref{S1} by the diamond geometry $D$ shown in \cref{fig:cd}, thereby defining a finite-lifetime worldsheet. From the perspective of an eternal Minkowski observer, this corresponds to a mapping $X^\mu(\tau,\sigma)\to \bar{X}^\mu(f(\tau,\sigma))$. Rather than specifying this mapping explicitly, we adopt a structurally guided approach, informed by analogous embeddings of Rindler and Milne geometries into string worldsheets \cite{Bagchi:2021ban,Bagchi:2020ats,Karan:2024jgn}.

The mapping $X^\mu(\tau,\sigma)\to \bar{X}^\mu(f(\tau,\sigma))$ is implemented through a set of structural assumptions. The causal restriction in the target space induces a corresponding causal structure on the worldsheet, endowing it with null horizons and degenerate features intrinsic to the diamond geometry. This is made explicit by a reparametrization of the coordinates $(\tau,\sigma)$ such that the worldsheet metric $\gamma_{ab}$ assumes the form of a two-dimensional causal diamond. This construction partitions the worldsheet into the diamond interior $D$ and its exterior $\bar{D}$ (see \cref{fig:cd}), defined by ${D} \coloneqq \left\lbrace (\tau,\sigma): |\tau| + |\sigma|/{c} \leq\upalpha\right\rbrace$ and $\bar{D} \coloneqq \left\lbrace (\tau,\sigma): |\tau|+ |\sigma|/{c} >\upalpha\right\rbrace$. The parameter $\upalpha$ defines the half-diamond size and sets the finite lifetime of the worldsheet, $\mathbb{T}=2\upalpha$. The limit $\upalpha \to \infty$ recovers the conventional Minkowski worldsheet and restores access to the full target space. In contrast, the limit $\upalpha = 0$ yields a degenerate configuration of the diamond worldsheet corresponding to its birth. As will be shown in \cref{limit}, this limit exposes the tensionless regime and induces a Carrollian structure characteristic of finite-lifetime strings.

It follows from the above construction that the diamond worldsheet $\bar{X}^\mu(f(\tau,\sigma))$ must be described in terms of new coordinates $\eta(\tau, \sigma)$ and $\xi(\tau, \sigma)$, and will henceforth be written as $\bar{X}^\mu(\eta, \xi)$. From the viewpoint of an observer on the Minkowski worldsheet, the following conformal transformations map $(\tau, \sigma) \rightarrow (\eta, \xi)$:
\begin{align}\label{S4}
\begin{gathered}
\frac{2\tau/\upalpha}{\left(1 - \sigma/c\upalpha\right)^2 - (\tau/\upalpha)^2} = \varepsilon e^{2\xi/c\upalpha}\sinh\left(2\eta/\upalpha\right), \\[3pt]
\frac{1 - (\sigma/c\upalpha)^2 + (\tau/\upalpha)^2}{\left(1 - \sigma/c\upalpha\right)^2 - (\tau/\upalpha)^2} = \varepsilon e^{2\xi/c\upalpha}\cosh\left(2\eta/\upalpha\right),
\end{gathered}
\end{align}
where $\varepsilon = \pm 1$ distinguishes the interior $D$ and exterior $\bar{D}$ regions of the diamond worldsheet. Notably, in expressing the Minkowski-to-diamond worldsheet coordinate transformations \eqref{S4}, we adopt the convention of \cite{Camblong:2024bsy}. Interested readers may find several alternative representations in the literature \cite{Martinetti:2002sz,Martinetti:2008ja,Ida:2012um,Su:2015oys,DeLorenzo:2017tgx,Jacobson:2018ahi,Chakraborty:2022qdr,Foo:2020rsy,Casini:2011kv}.

Inversely, the diamond worldsheet coordinates $(\eta, \xi)$ are defined via
\begin{align}\label{S5}
\begin{gathered}
\eta(\tau,\sigma) = \frac{\upalpha}{4} \ln \left[\frac{\sigma^2/c^2\upalpha^2 - (1 + \tau/\upalpha)^2}{\sigma^2/c^2\upalpha^2 - (1 - \tau/\upalpha)^2}\right], \\[3pt]
\xi(\tau,\sigma) = \frac{c\upalpha}{4} \ln \left[\frac{ (1 + \sigma/c\upalpha)^2 - \tau^2/\upalpha^2 }{(1 - \sigma/c\upalpha)^2 - \tau^2/\upalpha^2 }\right].
\end{gathered}
\end{align}
We further assume that the closed-string condition is preserved under the coordinate reparametrization \eqref{S5}. In the Minkowski worldsheet, closed strings satisfy the periodic boundary condition $X^\mu (\tau, \sigma) = X^\mu (\tau, \sigma + \ell)$. Under the transformation to diamond coordinates, the worldsheet $\bar{X}^\mu\left(\eta(\tau, \sigma), \xi(\tau, \sigma)\right)$ remains closed, but with a finite periodicity $\mathcal{L}_{\mathcal{D}}$ along the spatial $\xi$ direction,         
\begin{align}\label{S6}
 \begin{gathered}
 \bar{X}^\mu (\eta, \xi) = \bar{X}^\mu (\eta + \Delta\eta, \xi + \Delta\xi), \\[2pt]
\Delta\eta =  \eta(0,\ell) - \eta(0,0) = 0, \\ 
 \Delta\xi =  \xi(0,\ell) - \xi(0,0) = \frac{c\upalpha}{2} \ln \left(\frac{ 1 + \ell/c\upalpha }{1 - \ell/c\upalpha}\right) \equiv  \mathcal{L}_{\mathcal{D}}. 
\end{gathered}   
\end{align}

\subsection{Quantum mode expansion}\label{QME}

Since the metric structures in the Minkowski and diamond coordinate charts \eqref{S4} are related by conformal invariance, the corresponding worldsheets satisfy identical EOM:
\begin{align}\label{S8}
 (c^{-2} \partial_\eta^2 - \partial_\xi^2) \bar{X}^{\mu} = 0 =  (c^{-2} \partial_\tau^2 - \partial_\sigma^2) {X}^{\mu},
\end{align}
together with the Virasoro constraints, which in the $(\eta,\xi)$ parametrization take the form
\begin{align}\label{S9}
	\partial_\eta \bar{X}^\mu\partial_\xi \bar{X}_\mu = 0, \quad (c^{-1} \partial_\eta \bar{X}^\mu)^2 + (\partial_\xi \bar{X}^\mu)^2 = 0.
\end{align} 
It therefore follows that the quantum mode expansion on the diamond worldsheet can be constructed in direct analogy with the Minkowski case \eqref{S1}:
\begin{align}\label{S10}
\bar{X}^\mu (\eta, \xi)
&= \bar{x}^\mu + \alpha^\prime \bar{p}^\mu \eta + \sqrt{2\pi\alpha^\prime}\hspace{-0.13in}\sum_{\substack{n > 0;\\ \Lambda\in\{\rm D,\bar D\}}}\hspace{-0.13in}\Big[
\tensor*[^{\rm \Lambda}]{\tilde{\beta}}{_n^\mu}\tensor*[^{\rm \Lambda}]{\tilde{g}}{_n} + \tensor*[^{\rm \Lambda}]{{\beta}}{_n^\mu}\tensor*[^{\rm \Lambda}]{g}{_n} \nonumber \\
& \qquad\qquad\qquad\quad
 + \tensor*[^{\rm \Lambda}]{\tilde{\beta}}{_{-n}^\mu}\tensor*[^{\rm \Lambda}]{\tilde{g}}{_n^*}
+ \tensor*[^{\rm \Lambda}]{{\beta}}{_{-n}^\mu}\tensor*[^{\rm \Lambda}]{g}{_n^*}
\Big].
\end{align}
where $\Lambda$ labels the right- and left-moving diamond modes $\lbrace \tensor*[^{D,\bar D}]{\tilde{g}}{_n}, \tensor*[^{D,\bar D}]{{g}}{_n} \rbrace$ and their associated oscillators $\lbrace \tensor*[^{D,\bar D}]{\tilde{\beta}}{_n}, \tensor*[^{D,\bar D}]{{\beta}}{_n} \rbrace$ in the interior $D$ and exterior $\bar{D}$ regions. Notably, the oscillation operators are subject to the following commutation relations for all $\Lambda,\Lambda^\prime \in  \{D,\bar{D}\}$
\begin{align}\label{S19}
\big[\tensor*[^{\rm \Lambda}]{\tilde{\beta}}{_n^\mu},\tensor*[^{\rm {\Lambda^\prime}}]{\tilde{\beta}}{_m^\nu}\big] = \big[\tensor*[^{\rm \Lambda}]{\beta}{_n^\mu},\tensor*[^{\rm {\Lambda^\prime}}]{\beta}{_m^\nu}\big] = n\eta^{\mu\nu}\delta_{\Lambda,\Lambda^\prime}\delta_{n+m,0},
\end{align} 
and are related via their respective Hermitian conjugates $\tensor*[^{\rm \Lambda}]{\tilde{\beta}}{_{-n}^\mu}=\big(\tensor*[^{\rm \Lambda}]{\tilde{\beta}}{_n^\mu}\big)^\dagger$ and $\tensor*[^{\rm \Lambda}]{{\beta}}{_{-n}^\mu}=\big(\tensor*[^{\rm \Lambda}]{{\beta}}{_n^\mu}\big)^\dagger$. Furthermore, they annihilate the diamond worldsheet vacuum state $\lvert0_{\mathcal{D}}\rangle$, defined to obey:
\begin{align}\label{S20}
\tensor*[^{ D}]{\tilde{\beta}}{_n^\mu}\lvert0_{\mathcal{D}}\rangle = \tensor*[^{ \bar{D}}]{\tilde{\beta}}{_n^\mu}\lvert0_{\mathcal{D}}\rangle = \tensor*[^{D}]{{\beta}}{_n^\mu}\lvert0_{\mathcal{D}}\rangle = \tensor*[^{\bar{D}}]{{\beta}}{_n^\mu}\lvert0_{\mathcal{D}}\rangle = 0.
\end{align} 
Turning to the structure of the diamond worldsheet modes, we emphasize that they cannot be obtained by a naive substitution of $(\tau,\sigma)$ with $(\eta,\xi)$ in \eqref{S2.2}. Instead, they must satisfy the criterion that induces the causal diamond horizons separating the regions ${D}$ and $\bar{D}$ at the Minkowski level. Although the coordinate transformations \eqref{S4} and \eqref{S5} are algebraically involved, a more transparent formulation emerges upon introducing light-cone coordinates defined as
\begin{align}\label{S11}
U_{\rho} = c\tau + \rho \sigma, \quad  V_{\rho,\varepsilon} = \varepsilon(c\eta + \rho \xi),
\end{align}
where $\rho = \pm 1$ distinguishes the left- and right-moving modes. Here, the presence of $\varepsilon = \pm 1$ ensures that the diamond null coordinates are future-directed. The set \eqref{S11} therefore comprises the advanced and retarded Minkowski null coordinates $U_+ = U = c\tau + \sigma$ and $U_- = \tilde{U} = c\tau - \sigma$, together with their diamond counterparts $V_{+,+} = V_D = c\eta + \xi$ and $V_{-,+} = \tilde{V}_D = c\eta - \xi$ in the interior region $D$, and $V_{+,-} = V_{\bar{D}} = -c\eta - \xi$ and $V_{-,-} = \tilde{V}_{\bar{D}} = -c\eta + \xi$ in the exterior region $\bar{D}$. Remarkably, with these definitions, the transformation equations admit the following alternative form:
\begin{align}\label{S12}
\begin{gathered}
e^{{2V_D}/{c\upalpha}}= \frac{1 + {U}/{c\upalpha} }{1 - {U}/{c\upalpha} }, \quad   e^{-2\tilde{V}_D/{c\upalpha}}= \frac{1 - \tilde{U}/{c\upalpha} }{1 + \tilde{U}/{c\upalpha} },\\
e^{-{2V_{\bar{D}}}/{c\upalpha}}= \frac{{U}/{c\upalpha} + 1 }{ {U}/{c\upalpha} - 1 }, \quad   e^{2\tilde{V}_{{\bar{D}}}/{c\upalpha}}= \frac{ \tilde{U}/{c\upalpha} - 1 }{ \tilde{U}/{c\upalpha} + 1}.
\end{gathered}
\end{align}
At this stage, the diamond worldsheet modes are introduced in their most general form as
\begin{align}\label{S13}
g_n (V_{\rho,\varepsilon}) = {i  \over \sqrt{4\pi}n}e^{-i\Omega_n V_{\rho,\varepsilon}/c},
\end{align}
where they are associated with a redefined frequency $\Omega_{n}$, reflecting the reparametrized closed-string periodicity $\mathcal{L}_{\mathcal{D}}$ on the diamond worldsheet as defined in \eqref{S6},
\begin{align}\label{S7}
	\Omega_{n} = {2\pi cn \over \mathcal{L}_{\mathcal{D}} } = \frac{4\pi n}{\upalpha \ln \left(\frac{ c\upalpha + \ell }{c\upalpha - \ell }\right)}.
\end{align}
These mode functions may then be expressed in terms of the Minkowski null coordinates $U_\rho$ by mapping $V_{\rho,\varepsilon}$ accordingly, with support restricted to the interior region ${D}$ (via $U_{\rho}<\upalpha$) and the exterior region $\bar{D}$ (via $U_{\rho}>\upalpha$), yielding
\begin{align}\label{S14}
 g_n (U_\rho) = {i  \over \sqrt{4\pi}n}e^{-i\Omega_n V_{\rho,\varepsilon}(U_\rho)/c}H \big(\varepsilon (\upalpha - |U_\rho| )\big),
\end{align}
where $H(z)$ denotes the Heaviside step function. Using the mappings \eqref{S12}, the explicit right- and left-moving worldsheet mode functions in the interior region $D$ take the form
\begin{align}\label{S15}
 \tensor*[^{ D}]{{g}}{_n}(\tilde{U}) &= \tensor*[^{ D}]{\tilde{g}}{_n} = {i  \over \sqrt{4\pi}n}e^{-i\Omega_n \tilde{V}_{D}/c}H\big( \upalpha - |\tilde{U}|\big) \nonumber \\
 &= {i  \over \sqrt{4\pi}n} \left(\frac{1 - \tilde{U}/{c\upalpha} }{1 + \tilde{U}/{c\upalpha} }\right)^{i\upalpha\Omega_n/2}H\big( \upalpha - |\tilde{U}|\big),
\end{align}
and
\begin{align}\label{S16}
  \tensor*[^{ D}]{{g}}{_n}({U}) &= \tensor*[^{ D}]{{g}}{_n} = {i  \over \sqrt{4\pi}n}e^{-i\Omega_n {V}_{D}/c}H\big( \upalpha - |{U}|\big) \nonumber \\
 &= {i  \over \sqrt{4\pi}n} \left(\frac{1 + {U}/{c\upalpha} }{1 - {U}/{c\upalpha} }\right)^{-i\upalpha\Omega_n/2}H\big( \upalpha - |{U}|\big).
\end{align}
Similarly, the corresponding worldsheet mode functions in the exterior region $\bar{D}$ are
\begin{align}\label{S17}
 \tensor*[^{\bar D}]{{g}}{_n}(\tilde{U}) &= \tensor*[^{ \bar{D}}]{\tilde{g}}{_n} = {i  \over \sqrt{4\pi}n}e^{-i\Omega_n \tilde{V}_{\bar{D}}/c}H\big(|\tilde{U}| - \upalpha\big) \nonumber \\
 &= {i  \over \sqrt{4\pi}n} \left(\frac{ \tilde{U}/{c\upalpha} - 1 }{ \tilde{U}/{c\upalpha} + 1}\right)^{-i\upalpha\Omega_n/2}H\big(|\tilde{U}| - \upalpha \big),
\end{align}
and
\begin{align}\label{S18}
 \tensor*[^{\bar D}]{{g}}{_n}({U}) &= \tensor*[^{ \bar{D}}]{{g}}{_n} = {i  \over \sqrt{4\pi}n}e^{-i\Omega_n {V}_{\bar{D}}/c}H\big(|{U}| - \upalpha \big) \nonumber \\
 &= {i  \over \sqrt{4\pi}n} \left(\frac{ {U}/{c\upalpha} + 1 }{{U}/{c\upalpha} - 1}\right)^{i\upalpha\Omega_n/2}H\big(|{U}| - \upalpha \big).
\end{align}
It is worth noting that the corresponding conjugate modes appearing in the expansion \eqref{S10} are obtained by complex conjugation. These diamond worldsheet modes, expressed in Minkowski coordinates, form the basis for constructing the Unruh-type mode decomposition in the subsequent subsection.

\subsection{Unruh-diamond modes and Bogoliubov transformations}\label{Unruh}

In this subsection, we develop a crucial extension of the analytic continuation method originally introduced by Unruh in the context of Rindler spacetime \cite{Unruh:1976db}. The same strategy can be adapted to the diamond worldsheet by appropriately extending the local diamond modes in \cref{S15,S16,S17,S18} while retaining only their positive-frequency components. This procedure gives rise to what we refer to as the ``Unruh–diamond modes.'' These modes allow the diamond worldsheet to be analyzed from the viewpoint of a Minkowski observer through the Bogoliubov transformations that relate their respective quantum modes and oscillator operators. In this sense, the construction closely parallels the familiar derivation of Unruh modes in Rindler spacetime, now realized in the context of string worldsheets in a causal diamond background.

This construction is necessary because the local diamond modes are non-analytic and have support restricted to either $D$ or $\bar D$, i.e., $\tensor*[^{\rm D}]{\tilde{g}}{_n} = \tensor*[^{ D}]{{g}}{_n} = 0$ in $\bar D$, while $\tensor*[^{ \bar D}]{\tilde{g}}{_n} = \tensor*[^{ \bar D}]{{g}}{_n} = 0$ in $D$. They therefore do not define a global mode basis. To construct a global description, one forms specific linear combinations that admit analytic continuation across the diamond horizons, yielding a globally well-defined basis analogous to Unruh modes in Rindler spacetime. These combinations thus define the global mode basis of the diamond worldsheet from the perspective of a Minkowski worldsheet observer.

Technically, the procedure begins by identifying those modes of the diamond worldsheet that admit analytic continuation across the full spacetime, encompassing both the $D$ and $\bar D$ regions. However, the functional forms of the diamond modes involved (see \cref{S15,S16,S17,S18}) are multi-valued functions, and a specific branch prescription must therefore be chosen. Explicitly, one constructs suitable linear combinations of the local diamond modes whose analytic continuation in the complex $U_\pm$ plane ensures their positive-frequency character with respect to the Minkowski vacuum \eqref{S3}. The resulting Unruh–diamond modes thus provide a globally analytic basis from which the Bogoliubov relations between the Minkowski and diamond oscillator operators can be systematically derived.

Let us first construct the right-moving branch of the Unruh–diamond basis. Remarkably, the local mode functions \eqref{S15} and \eqref{S17} in region $D$, when appropriately paired with their Hermitian partners in $\bar D$, give rise to the following linear combinations:
\begin{align}
\tensor*[^{D}]{\tilde{g}}{_n} - (-1)^{\frac{i \Omega_{n}\upalpha} {2}}\,\tensor*[^{\bar D}]{\tilde{g}}{_n^*} &= {i \over \sqrt{\pi}n}\left(\frac{1 - \tilde{U}/{c\upalpha} }{1 + \tilde{U}/{c\upalpha} }\right)^{\frac{i\upalpha\Omega_n}{2}}\label{U1},\\
\tensor*[^{\bar D}]{\tilde{g}}{_n} - (-1)^{-\frac{i \Omega_{n}\upalpha}{2}}\,\tensor*[^{ D}]{\tilde{g}}{_n^*} &= {i \over \sqrt{\pi}n} \left(\frac{ \tilde{U}/{c\upalpha} - 1 }{ \tilde{U}/{c\upalpha} + 1}\right)^{-\frac{i\upalpha\Omega_n}{2}}.\label{U2}
\end{align}
These combinations are well defined and admit analytic continuation between $D$ and $\bar D$, thereby producing a set of right-moving global worldsheet modes. It is evident that the pairing \eqref{U1} performs a rotation between the two involved modes that projects the cumulative physics into the interior $D$ diamond worldsheet region, whereas in \eqref{U2} the corresponding structure maps into the exterior $\bar D$ region. Similarly, for the left-moving branch, we obtain the following analytically continued combinations:
\begin{align}
\tensor*[^{D}]{{g}}{_n} - (-1)^{-\frac{i \Omega_{n}\upalpha} {2}}\,\tensor*[^{\bar D}]{{g}}{_n^*} &= {i \over \sqrt{\pi}n}\left(\frac{1 + {U}/{c\upalpha} }{1 - {U}/{c\upalpha} }\right)^{-\frac{i\upalpha\Omega_n}{2}}, \label{U3}\\
\tensor*[^{\bar D}]{{g}}{_n} - (-1)^{\frac{i \Omega_{n}\upalpha} {2}}\,\tensor*[^{ D}]{{g}}{_n^*} &= {i \over \sqrt{\pi}n} \left(\frac{ {U}/{c\upalpha} + 1 }{{U}/{c\upalpha} - 1}\right)^{\frac{i\upalpha\Omega_n}{2}}. \label{U4}
\end{align}
One observes that two possible analytic continuations arise from the double signs in the relation $\log(-1)=\pm i\pi$, which enter all the combinations \eqref{U1}–\eqref{U4}. However, this ambiguity must be resolved by selecting the specific branch that yields a positive-frequency mode function. Only after this step can the local mode combinations, analytically continued across $U_\pm = 0$, be promoted to two types of global basis functions of the diamond worldsheet, consisting of the left- and right-moving sectors. Firstly, while projecting from $\bar D \to D$ we obtain,
\begin{align}\label{U5}
\begin{gathered}
 \tensor*[^{(1)}]{{h}}{_n}(U_\pm) = \frac{1}{\sqrt{1 - e^{-\pi\upalpha\Omega_n}}} \left[\tensor*[^{D}]{{g}}{_n}(U_\pm) - e^{-\frac{\pi \Omega_{n}\upalpha} {2}}\tensor*[^{\bar D}]{{g}}{_n^*}(U_\pm) \right], 
\end{gathered}
\end{align}
and secondly, while projecting from $D \to \bar D$
\begin{align}\label{U6}
\tensor*[^{(2)}]{{h}}{_n}(U_\pm) &= \frac{1}{\sqrt{1 - e^{-\pi\upalpha\Omega_n}}} \left[\tensor*[^{\bar D}]{{g}}{_n}(U_\pm) - e^{-\frac{\pi \Omega_{n}\upalpha} {2}}\tensor*[^{ D}]{{g}}{_n^*}(U_\pm) \right],
\end{align}
where the normalization factor is introduced so that the modes \eqref{U5} and \eqref{U6} form an orthonormal basis. For later convenience, we further parametrize these linear combinations using the variable
\begin{align}\label{U7}
\theta_n = \tanh{^{-1}}\left(-e^{-{\pi \upalpha\Omega_{n}\over 2}}\right),
\end{align}
which generates the following transformations between the global and local diamond worldsheet modes
\begin{align}\label{U8}
\begin{split}
\tensor*[^{(1)}]{{h}}{_n}(U_\pm) &= \cosh \theta_n \tensor*[^{ D}]{{g}}{_n}(U_\pm) + \sinh\theta_n \tensor*[^{\bar D}]{{g}}{_n^*}(U_\pm), \\[2pt]
\tensor*[^{(2)}]{{h}}{_n}(U_\pm) &= \cosh \theta_n \tensor*[^{\bar  D}]{{g}}{_n}(U_\pm) + \sinh\theta_n \tensor*[^{ D}]{{g}}{_n^*}(U_\pm).   
\end{split}
\end{align}
These relations explicitly exhibit the Bogoliubov-type mixing between the interior and exterior diamond worldsheet modes. The structure of this mixing differs from the conventional Unruh construction \cite{Unruh:1976db} in field theory, although a similar structure arises through analyticity across the causal diamond horizon \cite{Ida:2012um,Camblong:2024bsy}. In the present case, we show that the same structure persists at the level of the worldsheet theory, where a right- (or left-) moving global mode is assembled solely from right- (or left-) moving diamond worldsheet modes originating from both regions $D$ and $\bar D$. Unlike the Rindler or Milne worldsheet constructions \cite{Bagchi:2021ban,Bagchi:2020ats,Karan:2024jgn}, this mixing preserves chirality, with right- and left-moving sectors remaining decoupled. Thus, analytic continuation across the diamond horizon reorganizes the modes between the interior and exterior regions while preserving the chiral decomposition of the diamond worldsheet dynamics. This feature reflects the causal organization of the diamond geometry and provides a distinct realization of the Unruh construction for fields or string worldsheets associated with observers of finite lifetime.

In what follows, although the modes \eqref{U5} and \eqref{U6} differ from the local diamond modes $\lbrace \tensor*[^{ D}]{{g}}{_n}(U_\pm), \tensor*[^{\bar D}]{{g}}{_n}(U_\pm) \rbrace$, they share the same analytic structure as the Minkowski modes \eqref{S2.2}. As a result, they are naturally associated with the same Minkowski worldsheet vacuum state $\lvert 0_{\rm M}\rangle$. It is therefore convenient to expand the worldsheet directly in this Unruh–diamond basis. Accordingly, instead of mapping with the expansion \eqref{S2.1}, the diamond worldsheet can be written in terms of $\tensor*[^{(1,2)}]{{h}}{_n}(U_\pm) = \lbrace \tensor*[^{(1,2)}]{{h}}{_n},\tensor*[^{(1,2)}]{\tilde{h}}{_n} \rbrace$,
\begin{align}\label{U9}
\bar{X}^\mu (\eta, \xi) &= \bar{x}^\mu + \alpha^\prime \bar{p}^\mu\eta + \sqrt{2\pi\alpha^\prime} \hspace{-0.03in}\sum_{\substack{n > 0; \\ i = 1,2}} \hspace{-0.03in}\Big[ \tensor*[^{ (i)}]{\tilde{\gamma}}{_n^\mu} \tensor*[^{ (i)}]{\tilde{h}}{_n} \nonumber \\
& \quad +  \tensor*[^{(i)}]{{\gamma}}{_n^\mu}\tensor*[^{(i)}]{h}{_n} + \tensor*[^{(i)}]{\tilde{\gamma}}{_{-n}^\mu} \tensor*[^{(i)}]{\tilde{h}}{_n^*} + \tensor*[^{(i)}]{{\gamma}}{_{-n}^\mu} \tensor*[^{(i)}]{h}{_n^*}\Big],
\end{align} 
where now
\begin{align}\label{U10}
\tensor*[^{(1)}]{\tilde{\gamma}}{_n^\mu}\lvert0_{\rm M}\rangle = \tensor*[^{(2)}]{\tilde{\gamma}}{_n^\mu}\lvert0_{\rm M}\rangle = \tensor*[^{(1)}]{{\gamma}}{_n^\mu}\lvert0_{\rm M}\rangle = \tensor*[^{(2)}]{{\gamma}}{_n^\mu}\lvert0_{\rm M}\rangle = 0.    
\end{align}
Therefore, we can now relate the diamond operators $\lbrace \tensor*[^{D,\bar D}]{\tilde{\beta}}{_n}, \tensor*[^{D,\bar D}]{{\beta}}{_n} \rbrace$ with their global counterparts $\lbrace \tensor*[^{(1,2)}]{\tilde{\gamma}}{_n}, \tensor*[^{(1,2)}]{{\gamma}}{_n} \rbrace$ by first taking the following inner products from the expansion \eqref{S10}, 
\begin{align}\label{U11}
\tensor*[^{\rm \Lambda}]{\tilde{\beta}}{_n^\mu} = \left\langle \bar{X}^\mu,\tensor*[^{\rm \Lambda}]{\tilde{g}}{_n}\right\rangle, \enspace \tensor*[^{\rm \Lambda}]{{\beta}}{_n^\mu} = \left\langle \bar{X}^\mu,\tensor*[^{\rm \Lambda}]{{g}}{_n}\right\rangle, \enspace \forall~ \Lambda =  D,\bar D
\end{align}
followed by expanding $\bar{X}^\mu (\eta, \xi)$ in the form \eqref{U9}. These inner products are defined on the worldsheet in close analogy with the standard Klein–Gordon inner product in quantum field theory \cite{Birrell:1982ix}, and therefore satisfy
\begin{align}\label{U11.1}
\begin{gathered}
 \big\langle \tensor*[^{\Lambda}]{{g}}{_m}(U_\pm),\tensor*[^{\Lambda^\prime}]{{g}}{_n}(U_\pm)\big\rangle =  \delta_{\Lambda,\Lambda^\prime}\delta_{m,n},\\[2pt]
 \big\langle \tensor*[^{\Lambda}]{{g}}{_m^*}(U_\pm),\tensor*[^{\Lambda^\prime}]{{g}}{_n^*}(U_\pm)\big\rangle = - \delta_{\Lambda,\Lambda^\prime}\delta_{m,n}, \\[2pt]
 \big\langle \tensor*[^{\Lambda}]{{g}}{_m}(U_\pm), \tensor*[^{\Lambda^\prime}]{{g}}{_n^*}(U_\pm)\big\rangle =0,
\end{gathered}  
\end{align}
for all $\Lambda,\Lambda^\prime =  (D,\bar{D})$. Strictly speaking, the relation \eqref{U11} determines the annihilation operators, while the corresponding creation operators follow from Hermitian conjugation, $\tensor*[^{\rm \Lambda}]{\tilde{\beta}}{_{-n}^\mu}=\big(\tensor*[^{\rm \Lambda}]{\tilde{\beta}}{_n^\mu}\big)^\dagger$ and $\tensor*[^{\rm \Lambda}]{{\beta}}{_{-n}^\mu}=\big(\tensor*[^{\rm \Lambda}]{{\beta}}{_n^\mu}\big)^\dagger$. This procedure leads directly to the Bogoliubov transformations relating the local diamond operators to the global oscillators,
\begin{align}\label{U12}
\begin{split}
\tensor*[^{D,\bar D}]{\tilde{\beta}}{_n^\mu} &= \beta_+  \tensor*[^{(1, 2)}]{\tilde{\gamma}}{_n^\mu} + \beta_- \tensor*[^{(2, 1)}]{\tilde{\gamma}}{_{-n}^\mu}, \\
\tensor*[^{D,\bar D}]{{\beta}}{_n^\mu} &= \beta_+ \tensor*[^{(1, 2)}]{{\gamma}}{_n^\mu} + \beta_- \tensor*[^{(2, 1)}]{{\gamma}}{_{-n}^\mu},\\
\tensor*[^{D,\bar D}]{\tilde{\beta}}{_{-n}^\mu} &= \beta_+ \tensor*[^{(1, 2)}]{\tilde{\gamma}}{_{-n}^\mu} + \beta_- \tensor*[^{(2, 1)}]{\tilde{\gamma}}{_{n}^\mu}, \\
\tensor*[^{D,\bar D}]{{\beta}}{_{-n}^\mu} &= \beta_+ \tensor*[^{(1, 2)}]{{\gamma}}{_{-n}^\mu} + \beta_- \tensor*[^{(2, 1)}]{{\gamma}}{_{n}^\mu},
\end{split}
\end{align}
with the Bogoliubov coefficients given by
\begin{align}\label{U13}
\begin{gathered}
\beta_+ = \cosh\theta_n = {1 \over \sqrt{1 -e^{-\pi \upalpha\Omega_{n}}}}, \\
\beta_- = \sinh\theta_n = -{e^{-{\pi \upalpha\Omega_{n}\over 2}} \over \sqrt{1 -e^{-\pi \upalpha\Omega_{n}}}}.
\end{gathered}
\end{align}
These coefficients satisfy the canonical Bogoliubov identity \( \beta_+^2 - \beta_-^2 = 1 \), ensuring that the transformation preserves the oscillator commutation relations \eqref{S19}. Moreover, it follows from the operator relations \eqref{U12} that the diamond worldsheet vacuum \eqref{S20} can be written as a coherent state in terms of the Minkowski worldsheet vacuum $\lvert 0_{\rm M}\rangle$ and the global $\gamma$-operators \cite{Wald:1995yp},
\begin{align}\label{U13.1}
\lvert 0_{\mathcal{D}}\rangle &= \mathbb{Z}^{1/2} \prod_{n=1}^{\infty} \exp\!\bigg[-\frac{\tanh \theta_n}{n} \nonumber \\
&\qquad \times \left(\tensor*[^{(1)}]{\tilde{\gamma}}{_{-n}^\mu}\tensor*[^{(2)}]{\tilde{\gamma}}{_{-n}^\mu} + \tensor*[^{(1)}]{\gamma}{_{-n}^\mu}\tensor*[^{(2)}]{\gamma}{_{-n}^\mu}\right)\bigg]\lvert 0_{\rm M}\rangle.
\end{align}
where $\mathbb{Z} = \left({1 - e^{-\pi \upalpha\Omega_{n}}}\right)^{-1/2}$. In other words, to an observer in the Minkowski vacuum $\lvert 0_{\rm M}\rangle$, the state $\lvert 0_{\mathcal{D}}\rangle$ appears as a highly excited and entangled two-mode squeezed state, forming an evolving vacua parametrized by the lifetime $\upalpha$ of the diamond worldsheet. 

For later convenience, it is now useful to rewrite the relations in \eqref{U12} in a form that makes the correspondence with the standard Minkowski oscillators explicit. While \eqref{U12} provides the operator-level mapping $\big(\tensor*[^{D,\bar D}]{\tilde{\beta}}{_n^\mu}, \tensor*[^{D,\bar D}]{{\beta}}{_n^\mu}\big) \to \big( \tensor*[^{(1, 2)}]{\tilde{\gamma}}{_n^\mu}, \tensor*[^{(1, 2)}]{{\gamma}}{_n^\mu}\big)$ between the local diamond basis and the Unruh–diamond basis, our goal is to express the corresponding relation directly in terms of the Minkowski oscillators $(\tilde{\alpha},\alpha)$. The key observation is that the Unruh–diamond oscillators span the same positive-frequency sector as the Minkowski modes and hence define the same Minkowski vacuum. Consequently, one may pass from the Unruh–diamond basis to an equivalent Minkowski basis by taking suitable linear combinations of the $\gamma$-operators, or equivalently by choosing a basis in which the Unruh–diamond oscillators align with the Minkowski oscillators up to an overall normalization. In the present context, this recombination is implemented as $\tensor*[^{(1)}]{\tilde{\gamma}}{_{n}^\mu} + \tensor*[^{(2)}]{\tilde{\gamma}}{_{n}^\mu} \sim {\tilde{\alpha}}{_n^\mu}$ and $\tensor*[^{(1)}]{{\gamma}}{_{n}^\mu} + \tensor*[^{(2)}]{{\gamma}}{_{n}^\mu} \sim {{\alpha}}{_n^\mu}$. This, in turn, motivates analogous recombinations for the diamond oscillators, defined by ${\tilde{\beta}}{_n^\mu} = \tensor*[^{D}]{\tilde{\beta}}{_{n}^\mu} + \tensor*[^{\bar D}]{\tilde{\beta}}{_{n}^\mu}$ and ${{\beta}}{_n^\mu} = \tensor*[^{D}]{{\beta}}{_{n}^\mu} + \tensor*[^{\bar D}]{{\beta}}{_{n}^\mu}$. The corresponding creation operators follow from Hermitian conjugation. The eight operator relations obtained in \eqref{U12} across $D \leftrightarrow \bar D$ can therefore be recast into four equivalent global relations
\allowdisplaybreaks{
\begin{align}\label{BGV}
\begin{gathered}
{\tilde{\beta}}{_n^\mu} = {1 \over \sqrt{2\sinh\left(\frac{\pi \upalpha\Omega_{n}}{2}\right)}} \left[e^{{\pi \upalpha\Omega_{n}\over 4}}{\tilde{\alpha}}{_n^\mu} - e^{-{\pi \upalpha\Omega_{n}\over 4}} {{\alpha}}{_{-n}^\mu} \right], \\
{{\beta}}{_n^\mu} = {1 \over \sqrt{2\sinh\left({\pi \upalpha\Omega_{n}}\over {2}\right)}} \left[ e^{{\pi \upalpha\Omega_{n}\over 4}}{{\alpha}}{_n^\mu} - e^{-{\pi \upalpha\Omega_{n}\over 4}}{\tilde{\alpha}}{_{-n}^\mu} \right], \\
{\tilde{\beta}}{_{-n}^\mu} = {1 \over \sqrt{2\sinh\left({\pi \upalpha\Omega_{n}}\over {2}\right)}} \left[e^{{\pi \upalpha\Omega_{n}\over 4}}{\tilde{\alpha}}{_{-n}^\mu} - e^{-{\pi \upalpha\Omega_{n}\over 4}} {{\alpha}}{_{n}^\mu} \right],  \\
{{\beta}}{_{-n}^\mu} = {1 \over \sqrt{2\sinh\left({\pi \upalpha\Omega_{n}}\over {2}\right)}} \left[ e^{{\pi \upalpha\Omega_{n}\over 4}}{{\alpha}}{_{-n}^\mu} - e^{-{\pi \upalpha\Omega_{n}\over 4}}{\tilde{\alpha}}{_{n}^\mu} \right].
\end{gathered}
\end{align}}
These results provide the Bogoliubov mapping between the recombined form of the diamond operators and the Minkowski oscillators $(\tilde{\beta}, \beta) \to (\tilde{\alpha}, \alpha)$, which will be used in the analysis of the tensionless limit in \cref{limit}. In this framework, the evolving vacua structure \eqref{U13.1} can be reformulated in terms of the Minkowski oscillators as
\begin{align}\label{U18}
\lvert 0_{\mathcal{D}}\rangle 
&= \mathbb{Z}^{1/2} \prod_{n=1}^{\infty} \exp\left[\frac{e^{-{\pi \upalpha\Omega_{n}\over 2}}} {n} \,\tensor*{\tilde{\alpha}}{_{-n}}\cdot\tensor*{{\alpha}}{_{-n}}\right]\lvert 0_{\rm M}\rangle,  
\end{align}
and we will be interested in examining the associated physics at the birth point $\upalpha = 0$ of the diamond worldsheet.


\section{At the birth of diamond worldsheet} \label{limit}

In this section, we investigate the physics that may emerge at the birth point of the causal diamond worldsheet introduced in \cref{sec1}. As discussed earlier, the diamond worldsheet is characterized by the parameter $\upalpha$, which determines the finite lifetime $\mathbb{T}=2\upalpha$ of the accessible worldsheet region. Consequently, the causal diamond describes a worldsheet geometry associated with observers of finite causal lifetime. As $\upalpha$ decreases, the accessible region shrinks, and in the extreme limit $\upalpha = 0$, the worldsheet collapses to its birth point, where both temporal and spatial extents degenerate into a fully degenerate configuration.

Having said that, it is also instructive to examine the behavior of the diamond worldsheet metric at the birth point $\upalpha =0$. To this end, we start with the two-dimensional Minkowski worldsheet metric $ds^2= dUd\tilde{U}$ and then consider the following reparametrization of the $(\tau,\sigma)$ coordinates into $(\eta,\xi)$ (via simplifying \eqref{S12} in the interior $D$ region)
\begin{align}\label{bl1}
U = c\upalpha \tanh\left({V_D \over {c\upalpha}}\right),\quad \tilde{U} = c\upalpha \tanh\left({\tilde{V}_D \over {c\upalpha}}\right) .
\end{align}
From these relations one obtains the structure of the induced metric on the diamond worldsheet as
\begin{align}\label{bl2}
ds^2 = \gamma_{ab} d\xi^a d\xi^b =
\Omega_\mathcal{D}^2(\eta,\xi;\upalpha)d{V}_Dd\tilde{V}_D,
\end{align}
where $\xi^{a} = (\eta,\xi)$. This shows that the diamond worldsheet geometry is conformal to its flat Minkowski counterpart, with all nontrivial structure encoded in the conformal factor
\begin{align}\label{bl3}
\Omega_\mathcal{D}(\eta,\xi;\upalpha) &= 4 \left[\cosh\left(\frac{2\eta}{\upalpha}\right)+
\cosh\left(\frac{2\xi}{c\upalpha}\right)\right]^{-2},
\end{align}
yielding the determinant of the induced metric $\det \gamma_{ab} =-\Omega_\mathcal{D}^4(\eta,\xi;\upalpha)$. The degenerate points are identified by the vanishing of this conformal factor \eqref{bl3}, which occurs as the diamond worldsheet approaches the limits
\begin{align}\label{bl4}
 \frac{\eta}{\upalpha} \to \pm\infty, \quad \frac{\xi}{c\upalpha} \to \pm\infty.
\end{align}
More strikingly, while these limits may appear to characterize horizon-type degeneracies, the case $\upalpha=0$ is qualitatively distinct. For fixed $(\eta,\xi)$, the arguments of the hyperbolic functions diverge as $\upalpha\to0$, causing the conformal factor in \eqref{bl3} to vanish almost everywhere on the worldsheet. This behavior demonstrates that the birth-point limit corresponds to a globally degenerate worldsheet configuration, where the conformal factor collapses and the notion of local propagation breaks down. This stands in sharp contrast to the usual notion of degeneracy in eternal worldsheets, which has so far remained local and confined to null horizons. Even in non-inertial worldsheet constructions, null degeneracies arise only in limiting regimes, such as the ultra-high frequency limit of their corresponding characteristic parameters \cite{Bagchi:2021ban,Karan:2024jgn}. In this regime, the worldsheet dynamics reorganize into an ultra-local structure, consistent with the emergence of a tensionless string phase characterized by Carrollian symmetry. Here, the degeneration is intrinsically global, signaling a transition to a distinct physical phase rather than a localized geometric feature.

It is noteworthy that null degeneration can also be realized within the diamond worldsheet under the Carrollian contraction $c\to0$ at fixed $\upalpha$, which suppresses the temporal component of the induced metric \eqref{bl2}. Thus, while for finite $\upalpha$ the degeneration remains restricted to the null boundaries of the diamond, at the birth point $\upalpha=0$ the entire worldsheet enters a globally degenerate phase. These two limits therefore represent inequivalent mechanisms leading to degeneration: one geometric (finite lifetime), the other kinematical (Carrollian contraction).


Nevertheless, the degenerate points of worldsheet geometries are known to play a special role in several previously explored eternal worldsheet constructions. In particular, it has been observed that near such degenerate configurations the effective dynamics may reorganize in a way that resembles the structure associated with tensionless or Carrollian strings \cite{Schild:1976vq,Isberg:1993av,Bagchi:2015nca,Cardona:2016ytk,Bagchi:2013bga} (also see the review \cite{Bagchi:2026wcu} and references therein). This naturally motivates the question of whether the birth point $\upalpha=0$ realizes such a reorganization in an intrinsic manner.

At present, no intrinsic framework exists for defining a tensionless structure for a finite-lifetime (diamond) worldsheet directly at $\upalpha = 0$. Attempting to construct the worldsheet modes at this stage would lead to a degenerate configuration where the usual quantum mode expansion and oscillator description become ill-defined. Thus, the only controlled approach is to probe this regime through a limiting procedure starting from the well-defined tensile construction in \cref{sec1}. Similar strategies have proven useful in several other non-inertial and eternal worldsheet settings. For example, tensionless Rindler (accelerated) and Milne (time-evolving) worldsheets have been explored by analyzing appropriate horizon limits of the corresponding tensile constructions \cite{Bagchi:2020ats,Bagchi:2021ban,Karan:2024jgn}.

Guided by these considerations, we now examine how the Bogoliubov transformations \eqref{BGV} behave as the diamond worldsheet approaches its birth point. Our aim is to analyze the structure of these relations in the $\upalpha = 0$ regime and determine whether their truncated form exhibits features reminiscent of the oscillator mappings known to arise in tensionless worldsheet dynamics. The analysis below therefore serves as a diagnostic step toward understanding the possible role of tensionless physics in the finite-lifetime diamond worldsheet framework.


\subsection{Truncations at global and local degenerate points}\label{truncation}

Technically speaking, we want to investigate whether the behaviors of the Bogoliubov mappings \eqref{BGV} remain consistent when expanded in the limits $\upalpha \to 0$ and $c \to 0$. In other words, we wish to examine whether the expected correspondence $\text{tensionless}\leftrightarrow\text{Carrollian strings}$ continues to hold on the diamond worldsheet. Carrying out these truncations separately would unnecessarily complicate the intermediate algebra and obscure the structure of the limit. A more natural approach is therefore to consider the combined limit $c\upalpha \to 0$, which simultaneously satisfies the degeneracy condition \eqref{bl4} and captures both the collapse of the causal domain toward its birth point and the Carrollian contraction of the diamond worldsheet geometry.

To expose the structure of this limiting truncation, we reorganize the diamond mode frequency parameter $\Omega_n$ introduced in \eqref{S7} and identify it as the dominant quantity controlling the transcendental factors appearing in \eqref{BGV}. In particular, the Bogoliubov transformations are governed by
\begin{align}\label{T1}
{\upalpha\Omega_n}
=\frac{4\pi n}{\ln\left(\frac{c\upalpha+\ell}{c\upalpha-\ell}\right)}.
\end{align}
The asymptotic structure is therefore controlled by the logarithm in the denominator, which must be handled carefully in order to realize the limit $c\upalpha \to 0$. Working in the regime $c\upalpha \ll \ell$, it is convenient to rewrite and expand the logarithm as
\begin{align}\label{T2}
\ln\left(\frac{c\upalpha+\ell}{c\upalpha-\ell}\right) &= si\pi+2\operatorname{arctanh}\left(\frac{c\upalpha}{\ell}\right) \nonumber\\
&= si\pi+\frac{2c\upalpha}{\ell}+ \mathcal{O}((c\upalpha)^3),
\quad s=\pm1.
\end{align}
Here the sign $s$ keeps track of the two admissible branches of $\ln(-1)=\pm i\pi$. Hence, the phase entering the Bogoliubov coefficients behaves as
\begin{align}\label{T3}
\frac{\pi\upalpha\Omega_n}{2} = -2s\pi in + \frac{4cn}{\ell}\upalpha + \mathcal{O}((c\upalpha)^3).
\end{align}
This is the crucial point of the truncation. The phase term $\frac{\pi\upalpha\Omega_n}{2}$ does not vanish as $c\upalpha \to 0$; rather, it approaches the finite imaginary branch value $\mp 2\pi i n$. The asymptotic structure is therefore governed by deviations from this branch value, which scale with $c\upalpha/\ell$. It is this deviation, rather than the full phase, that controls the Bogoliubov prefactor in this limit. In this sense, the limit $c\upalpha \to 0$ should be interpreted as a controlled expansion around the shifted phase $\mp 2\pi i n$. Following this prescription, the factors appearing in the Bogoliubov transformations are expanded as
\begin{align}
e^{\pm\frac{\pi\upalpha\Omega_n}{4}} &= e^{\mp s\pi in}\,
e^{\pm\frac{2cn}{\ell}\upalpha+\mathcal{O}((c\upalpha)^3)} \nonumber \\
&= (-1)^{n} \left( 1 \pm \frac{2cn}{\ell}\upalpha + \mathcal{O}((c\upalpha)^3) \right),\label{T4}
\end{align}
and
\begin{align}
\sinh\left(\tfrac{\pi\upalpha\Omega_n}{2}\right)
&= \sinh\left(-2s\pi in + \frac{4cn}{\ell}\upalpha + \mathcal{O}((c\upalpha)^3) \right) \nonumber \\ 
&= \frac{4cn}{\ell}\upalpha + \mathcal{O}((c\upalpha)^3). \label{T5} 
\end{align}
The complete Bogoliubov coefficients therefore take the asymptotic form
\begin{align}\label{T6}
\frac{e^{\pm\frac{\pi\upalpha\Omega_n}{4}}}
{\sqrt{2\sinh\!\left(\tfrac{\pi\upalpha\Omega_n}{2}\right)}} = \frac{(-1)^n}{2}\left(
\sqrt{\tfrac{\ell}{2cn\upalpha}} \pm \sqrt{\tfrac{2cn\upalpha}{\ell}} \right) + \mathcal{O}((c\upalpha)^{3/2}).
\end{align}
Evidently, the overall factor $(-1)^n$ originates from the branch structure of the logarithm and can be absorbed into a harmless redefinition of the oscillator modes.\footnote{The phase factor $(-1)^n$ can be absorbed by redefining the oscillators mode-by-mode, $\beta_n^\mu \rightarrow (-1)^n \beta_n^\mu$ and $\tilde{\beta}_n^\mu \rightarrow (-1)^n \tilde{\beta}_n^\mu$. The oscillator algebra, analogous to \eqref{S19}, remains unchanged since the commutator acquires a factor $(-1)^{n+m}$, which equals unity on the support of $\delta_{n+m,0}$.} We therefore suppress this factor in what follows. Substituting the asymptotic prefactor into the Bogoliubov relations \eqref{BGV} and retaining the leading contributions in $c\upalpha$, the oscillator mappings reorganize into the truncated form
\allowdisplaybreaks{
\begin{align}\label{T7X}
\begin{gathered}
 \tilde{\beta}_n^\mu = \frac12 \left[
\left(\sqrt{\tfrac{2cn\upalpha}{\ell}} + \sqrt{\tfrac{\ell}{2cn\upalpha}} \right)\tilde{\alpha}_n^\mu + \left(\sqrt{\tfrac{2cn\upalpha}{\ell}} - \sqrt{\tfrac{\ell}{2cn\upalpha}}\right)\alpha_{-n}^\mu \right], \\[1pt]
\beta_n^\mu = \frac12 \left[ \left(\sqrt{\tfrac{2cn\upalpha}{\ell}} - \sqrt{\tfrac{\ell}{2cn\upalpha}}\right)\tilde{\alpha}_{-n}^\mu +
\left(\sqrt{\tfrac{2cn\upalpha}{\ell}} + \sqrt{\tfrac{\ell}{2cn\upalpha}} \right){\alpha}_n^\mu  \right], \\[2pt]
 \tilde{\beta}_{-n}^\mu = \frac12 \left[
\left(\sqrt{\tfrac{2cn\upalpha}{\ell}} + \sqrt{\tfrac{\ell}{2cn\upalpha}} \right)\tilde{\alpha}_{-n}^\mu + \left(\sqrt{\tfrac{2cn\upalpha}{\ell}} - \sqrt{\tfrac{\ell}{2cn\upalpha}}\right)\alpha_{n}^\mu \right],\\[2pt]
\beta_{-n}^\mu = \frac12 \left[ \left(\sqrt{\tfrac{2cn\upalpha}{\ell}} - \sqrt{\tfrac{\ell}{2cn\upalpha}}\right)\tilde{\alpha}_{n}^\mu + 
\left(\sqrt{\tfrac{2cn\upalpha}{\ell}} + \sqrt{\tfrac{\ell}{2cn\upalpha}} \right){\alpha}_{-n}^\mu \right].  
\end{gathered} 
\end{align}}  
The divergence of these relations as $(c\upalpha)^{-1/2}$ signals the emergence of a highly excited squeezed vacuum structure, as already encoded in \eqref{U18}. This suggests that the birth-point limit of the diamond worldsheet naturally drives the theory toward a new non-trivial phase. In particular, inserting the same small-phase expansion \eqref{T3} into the vacuum expression \eqref{U18}, the diamond worldsheet vacuum takes the asymptotic form
\begin{align}\label{T12}
\lvert 0_{\mathcal{D}}\rangle 
&= \hat{\mathbb{Z}}^{1/2} \prod_{n=1}^{\infty}
\exp\left[
\left(
\frac{1}{n} - \frac{4c\upalpha}{\ell} + \mathcal{O}((c\upalpha)^2)
\right)
\tensor*{\tilde{\alpha}}{_{-n}}\cdot\tensor*{{\alpha}}{_{-n}}
\right]
\lvert 0_{\rm M}\rangle,
\end{align}
where $\hat{\mathbb{Z}}$ is an (infinite) normalization constant. Thus, in the strict limit $c\upalpha \to 0$, the vacuum behaves at leading order as
\begin{align}\label{T13}
\lvert 0_{\mathcal{D}}\rangle 
\sim \hat{\mathbb{Z}}^{1/2} \prod_{n=1}^{\infty} 
\exp\left[
\frac{1}{n}\,\tensor*{\tilde{\alpha}}{_{-n}}\cdot\tensor*{{\alpha}}{_{-n}}
\right]
\lvert 0_{\rm M}\rangle.
\end{align}
In the next subsection, we interpret this emergent phase transition at leading order in the $c\upalpha \to 0$ truncation and elucidate its underlying physical meaning in the dynamics of the diamond worldsheet.

\subsection{Emergence of a tensionless phase}\label{tensionless}

We now interpret the degenerate regime identified in the previous subsection and demonstrate that it signals the emergence of a tensionless phase on the diamond worldsheet. To make this connection precise, we begin by recalling a set of well-established structural criteria (\hyperref[I]{Class I}, \hyperref[II]{II}, \hyperref[III]{III}, and \hyperref[IV]{IV}) that any consistent realization of tensionless string dynamics is expected to satisfy. The dynamics of a tensionless closed-string worldsheet on a flat Minkowski background is governed by the ILST action \cite{Isberg:1993av}
\begin{align}\label{T14}
\mathcal{S}_{\rm ILST} = \int d\tau d\sigma\, V^aV^b\partial_aX^{\mu}\partial_bX^{\nu}\eta_{\mu\nu},
\end{align}
where $V^a$ is a vector density that replaces the role of the tensile worldsheet metric appearing in \eqref{S1} in the $\mathcal{T}\to 0$ limit. Working in the gauge $V^a=(1,0)$, the EOM/constraints become
\begin{align}\label{T15}
  {\partial_\tau}^2 X^\mu =0, \quad \left(\partial_\tau X^\mu\right)^2 = \left(\partial_\tau X^\mu\right)\left(\partial_\sigma X_\mu\right) = 0.
\end{align}
The most general solution to this EOM was derived in \cite{Bagchi:2015nca}. For our purposes, it is convenient to reformulate it in the following mode-expanded form
\begin{align}\label{T16}
X^\mu (\tau, \sigma) &= x^\mu + \hat{\alpha}^\prime \hat{p}^\mu\tau \nonumber \\
& + \sqrt{2\pi\hat{\alpha}^\prime}\sum_{n > 0} \left[ \mathbb{\tilde{C}}_n^\mu \tilde{\Psi}_n + \mathbb{C}_n^\mu \Psi_n + \mathbb{\tilde{C}}_{-n}^\mu \tilde{\Psi}_n^* + \mathbb{C}_{-n}^\mu \Psi_n^* \right],
\end{align}
along with the right- and left-moving modes defined by
\begin{align}\label{T17}
\begin{gathered}
\tilde{\Psi}_n = {i \over \sqrt{4\pi}n}\left(1-i\omega_n\tau\right)e^{i\omega_n \sigma/c}, \\ 
\Psi_n = {i \over \sqrt{4\pi}n}\left(1-i\omega_n\tau\right) e^{-i\omega_n \sigma/c},
\end{gathered}
\end{align}
where $\mathbb{\tilde{C}}_0^\mu = \mathbb{C}_0^\mu = \sqrt{\hat{\alpha}^\prime \over 2}\hat{p}^\mu$. The constant $\hat{\alpha}^\prime$ and the corresponding zero-mode sector thus play roles analogous to the parameter $\alpha^\prime$ (proportional to the inverse string tension) and the zero-mode contribution appearing in the tensile mode expansion \eqref{S2.1}. The oscillator basis $\{\mathbb{\tilde{C}}_n^\mu, \mathbb{C}_n^\mu\}$ introduced in \eqref{T16} behaves as two decoupled sets of quantum harmonic oscillators, satisfying the commutation relations
\begin{align}\label{T18}
\big[\mathbb{\tilde{C}}_n^\mu, \mathbb{\tilde{C}}_m^\nu\big] =  \big[\mathbb{C}_n^\mu, \mathbb{C}_m^\nu\big] = n\eta^{\mu\nu}\delta_{n+m,0},
\end{align}
and annihilating a vacuum $\lvert 0_{\mathbb{C}}\rangle$ that is distinct from the tensile Minkowski vacuum $\lvert 0_{\rm M}\rangle$,
\begin{align}\label{T19}
\lvert 0_{\mathbb{C}}\rangle :~	\tensor*{\mathbb{\tilde{C}}}{_n^\mu}\lvert 0_{\mathbb{C}}\rangle = \tensor*{\mathbb{C}}{_n^\mu}\lvert 0_{\mathbb{C}}\rangle  = 0.
\end{align} 

\paragraph{Class I}\label{I} A striking feature of tensionless string theory is that, in the vanishing-tension limit, strings become long, floppy, and effectively \textit{infinitely stretchable} \cite{Bagchi:2015nca,Bagchi:2013bga}. This behaviour admits a precise realization at the level of the tensile mode expansion \eqref{S1}, which accommodates the tensionless limit through a singular (degenerate) rescaling of the worldsheet time coordinate,
\begin{align}\label{T20}
\sigma \to \sigma, \quad \tau \to \epsilon \tau, \quad \alpha^\prime \equiv {\hat{\alpha}^\prime \over \epsilon} \qquad \epsilon \to 0.
\end{align}

\paragraph{Class II}\label{II} The limit \eqref{T20} simultaneously induces an infinite boost on the worldsheet, ${\beta} = {{\sigma} \over c\tau} \to \infty$. This is naturally realized as a \textit{Carrollian limit} \cite{Duval:2014uoa, deBoer:2021jej} of the underlying worldsheet dynamics,
\begin{align}\label{T21}
c \to \epsilon c, \quad {{\beta}}  \equiv {{\hat{\beta}} \over \epsilon} \qquad \epsilon \to 0,
\end{align}
where ${\hat{\beta}}$ denotes the Carroll boost parameter. The tensionless regime is thus intrinsically associated with an ultra-relativistic (Carrollian) structure on the worldsheet.

\paragraph{Class III}\label{III} A more refined characterization emerges by comparing the tensile mode expansion \eqref{S1} with the tensionless mode expansion \eqref{T16} in the small-$\epsilon$ regime. One finds that the two descriptions become equivalent, provided the corresponding oscillators are related through a set of \textit{Bogoliubov transformations} \cite{Bagchi:2015nca}
 \allowdisplaybreaks{
\begin{align}\label{T22X}
\begin{gathered}
\tilde{\mathbb{C}}_n^\mu = \frac{1}{2}\left(\sqrt{\epsilon} + \frac{1}{\sqrt{\epsilon}}\right) \tilde{\alpha}_n^\mu + \frac{1}{2}\left(\sqrt{\epsilon} - \frac{1}{\sqrt{\epsilon}}\right) {\alpha}_{-n}^\mu, \\[1pt]
{\mathbb{C}}_{-n}^\mu = \frac{1}{2}\left(\sqrt{\epsilon} - \frac{1}{\sqrt{\epsilon}}\right) \tilde{\alpha}_{n}^\mu + \frac{1}{2}\left(\sqrt{\epsilon} + \frac{1}{\sqrt{\epsilon}}\right) {\alpha}_{-n}^\mu, \\[1pt]
{\mathbb{C}}_n^\mu = \frac{1}{2}\left(\sqrt{\epsilon} - \frac{1}{\sqrt{\epsilon}}\right) \tilde{\alpha}_{-n}^\mu + \frac{1}{2}\left(\sqrt{\epsilon} + \frac{1}{\sqrt{\epsilon}}\right) {\alpha}_{n}^\mu, \\[1pt]
\tilde{\mathbb{C}}_{-n}^\mu = \frac{1}{2}\left(\sqrt{\epsilon} + \frac{1}{\sqrt{\epsilon}}\right) \tilde{\alpha}_{-n}^\mu + \frac{1}{2}\left(\sqrt{\epsilon} - \frac{1}{\sqrt{\epsilon}}\right) {\alpha}_{n}^\mu.     
\end{gathered}
\end{align}}

\paragraph{Class IV}\label{IV} Although the transformations \eqref{T22X} may appear to interpolate across the parameter range $\epsilon \in [1,0]$, their validity is restricted to the small-$\epsilon$ regime, and they become singular in the limit $\epsilon \to 0$. Remarkably, this singular behavior is interpreted as a \textit{transition from a closed-string vacuum to an open-string boundary state} \cite{Bagchi:2019cay}, characterized by \textit{Dirichlet boundary conditions} \cite{Polchinski:1998rq} in all directions:
\begin{align}\label{T26}
\begin{gathered}
\left(\tilde{\alpha}_n^\mu - {\alpha}_{-n}^\mu\right)\lvert 0_{\mathbb{C}}\rangle =0, \\ 
\lvert 0_{\mathbb{C}}\rangle = \mathcal{N}_{\rm D}\prod_{n=1}^{\infty}\exp\left({1\over n}\,\tensor*{\tilde{\alpha}}{_{-n}}\cdot\tensor*{{\alpha}}{_{-n}}\right)\lvert 0_{\rm M}\rangle,  
\end{gathered}
\end{align}
At this stage, one immediately identifies that the leading-order Bogoliubov relations \eqref{T7X}, together with the resulting vacuum structure \eqref{T13}, in the limit $c\upalpha\to0$, precisely reproduce the structures summarized in \hyperref[III]{Class III} and \hyperref[IV]{IV}. This demonstrates that the degeneration of the diamond worldsheet reorganizes the tensile closed-string basis into a highly squeezed sector with the same operator structure as the tensionless theory. In particular, this establishes that 
\begin{align}\label{T27}
\epsilon \equiv \frac{2nc\upalpha}{\ell} \to 0: \enspace \big\lbrace \tilde{\beta}_n^\mu, {\beta}_n^\mu \big\rbrace \to \big\lbrace \tilde{\mathbb{C}}_n^\mu, {\mathbb{C}}_n^\mu \big\rbrace, \enspace \lvert 0_{\mathcal{D}}\rangle \to \lvert 0_{\mathbb{C}}\rangle.
\end{align} 
This equivalence establishes that the tensionless phase on the diamond worldsheet emerges through the following interconnected geometric routes:
\begin{itemize}

\item[\textbf{1.}] $\upalpha \to 0$ (for fixed $c, \ell$): keeping the closed-string diamond profile fixed, the lifetime of the worldsheet shrinks to zero. The finite-lifetime domain then collapses to a degenerate configuration, so that the theory is driven toward its birth configuration. In this sense, the moment of birth of the diamond worldsheet naturally realizes the onset of the tensionless phase.  

\item[\textbf{2.}] $c \to 0$ (for fixed $\upalpha, \ell$): one may instead approach the horizon of the two-dimensional diamond worldsheet along constant-$\upalpha$ trajectories while preserving the closed-string periodicity. This drives the worldsheet into an ultra-relativistic regime and induces the Carrollian structure encoded in \hyperref[II]{Class II}, thereby realizing the tensionless limit from the horizon side. 

\item[\textbf{3.}] $\ell \to \infty$ (for fixed $\upalpha, c$): if the periodicity scale diverges while the lifetime and worldsheet velocity remain fixed, the closed diamond worldsheet is stretched over an arbitrarily large spatial extent. Degenerate points then emerge around which the closed string becomes infinitely elongated and effectively fills the accessible spacetime region, reproducing the infinitely floppy behavior characteristic of \hyperref[I]{Class I}.     

\end{itemize}
All three routes correspond to distinct geometric mechanisms; however, they are unified by the identification $\epsilon \cong \frac{c\upalpha}{\ell}$ and thereby realize the same tensionless string physics. Crucially, in the diamond worldsheet this phase is not confined to special loci but emerges as a global property at birth. In this sense, the degeneration of the diamond worldsheet geometry, originating from the finite lifetime of the string, implies that the tensionless phase is no longer merely a kinematic limit, but instead constitutes an intrinsic phase of the theory characterized by a global, ultra-local Carrollian structure. This marks a transition to a regime where worldsheet propagation becomes strongly constrained, driving the dynamics toward an ultra-local structure characteristic of tensionless strings, consistent with the emergence of Carrollian worldsheet behavior in the tensionless limit. The birth configuration of the diamond worldsheet is thus identified as a new intrinsic origin of a Carrollian string phase, arising from this global degeneracy. Unlike conventional constructions tied to null horizons, this phase originates from the collapse of the causal domain itself, revealing a fundamentally new geometric origin of Carrollian string dynamics.


\section{Concluding remarks and outlook} \label{diss}

In this work, we have initiated the study of finite-lifetime string dynamics and constructed their worldsheet description within a causal diamond, thereby providing the first explicit realization of tensionless string dynamics in a finite-lifetime worldsheet setting. A striking outcome of our analysis is that \textit{strings are tensionless at birth and become tensile as they age, remaining so until their eventual death}. 

A key technical advance of this work is the direct physical interpretation of the $\epsilon$-parametric evolution \eqref{T20} within the parent Minkowski framework. In contrast to foundational and follow-up approaches (e.g., \cite{Bagchi:2015nca,Bagchi:2013bga,Bagchi:2019cay,Bagchi:2024epw}) based on inertial worldsheets and assuming eternal ($\mathbb{T} \to \infty$) strings, where this parameter was treated as an abstract contraction tool to access tensionless limits at $\epsilon \to 0$, our construction demonstrates that $\epsilon$ admits a concrete geometric realization. In particular, it encodes the evolution of the string lifetime from birth ($\mathbb{T}=0$) to death ($\mathbb{T}=2\upalpha$), controlled by the diamond worldsheet with fixed half-size parameter $\upalpha$. Crucially, this construction establishes a direct mapping between two Minkowski worldsheet observers—one eternal and one with finite lifetime—within a Unruh-like framework \cite{Unruh:1976db} formulated within the present development of finite-lifetime string theory.

This is in the same spirit as earlier constructions where one had to go beyond the parent Minkowski framework in order to relate the $\epsilon$-parameter evolution toward the limit $\epsilon \to 0$ in terms of non-inertial worldsheets \cite{Bagchi:2020ats,Bagchi:2021ban,Karan:2024jgn}. In fact, our framework is sufficiently flexible to establish a conformal connection between the diamond half-size $\upalpha$ and the acceleration parameter $a$ characterizing a tensionless Rindler worldsheet \cite{Bagchi:2020ats,Bagchi:2021ban}. In particular, one can modify the coordinate relations \eqref{S12} using a dimensionless parameter $\delta = \tfrac{2c}{a\upalpha}$ such that the diamond worldsheet interior $D$ transforms into the right-wedge of a Rindler worldsheet via\footnote{For details, one may follow the steps outlined in \cite{Camblong:2024bsy} to verify how the setting \eqref{D1} promotes the diamond worldsheet coordinates into Rindler-like form, $\tau = \tfrac{c}{a}e^{a\xi/c^2}\sinh{\left(a\eta/c\right)},\, \sigma = \tfrac{c^2}{a}e^{a\xi/c^2}\cosh{\left(a\eta/c\right)}$.}
\begin{align}\label{D1}
\delta e^{{aV_D}/{c^2}}= \frac{1 + {U}/{c\upalpha} }{1 - {U}/{c\upalpha} }, \quad \delta e^{-a\tilde{V}_D/{c^2}}= \frac{1 - \tilde{U}/{c\upalpha} }{1 + \tilde{U}/{c\upalpha} }.
\end{align}
With the light-cone coordinates vanishing at the center (birth) of the diamond worldsheet, one can expand the relations \eqref{D1} and enforce their mutual compatibility. This yields $\delta \left(1 + \tfrac{aV_D}{c^2}\right) \sim 1 + \tfrac{2U}{c\upalpha}$ and $\delta \left(1 - \tfrac{a\tilde{V}_D}{c^2}\right) \sim 1 - \tfrac{2\tilde{U}}{c\upalpha}$, which imply
\begin{align}
\delta = 1, \quad a = \frac{2c}{\upalpha} \implies \epsilon \simeq \upalpha \propto \frac{1}{a} \to 0.
\end{align}
These relations fix the acceleration parameter $a$ in terms of $\upalpha$, ensuring the emergence of the tensionless phase in the ultra-high acceleration limit $a \to \infty$ \cite{Bagchi:2020ats,Bagchi:2021ban}.

This study shows that the tensionless phase of finite-lifetime strings is not externally imposed, but emerges intrinsically as a robust degenerate sector of the diamond worldsheet at its birth, where the degeneration of the geometry signals a reorganization of the underlying degrees of freedom. Importantly, this degeneration is global rather than horizon-local, distinguishing it sharply from conventional null-horizon constructions. This is consistent with, while extending, the interpretation of tensionless strings as degenerate worldsheets endowed with Carrollian structure at null horizons. Furthermore, the associated Bogoliubov reorganization \eqref{T27} of the vacuum and oscillator structure faithfully reproduces the characteristic features of tensionless strings, where the vacuum assumes a squeezed, Dirichlet boundary-like form marking a closed-to-open string state transition, and the degrees of freedom differ fundamentally from those of the tensile theory. Our construction thus provides a concrete and structurally well-defined step toward a classical and quantum description of finite-lifetime null or Carrollian strings, addressing a long-standing open problem. Importantly, this shows that Carrollian string dynamics need not be tied exclusively to null-horizon structures, but can arise intrinsically from finite-lifetime worldsheet geometries as a global, ultra-local phase at birth, thereby revealing a previously unexplored origin of Carrollian behavior in string theory.

The framework developed here opens several directions for further investigation. In the latest parlance of the tensionless limit, the two copies of the worldsheet Virasoro algebra are also replaced by the conformal Carrollian algebra, which is isomorphic to the $\mathrm{BMS}_3$ algebra \cite{Duval:2014uva,Bagchi:2013bga}. The present finite-lifetime setting provides a natural arena in which such a transition may be dynamically realized, and it is particularly interesting to understand how this aligns with the symmetry structure of the diamond worldsheet, especially near the birth moment where the tensionless phase emerges. If successful, this would identify the birth point of the diamond worldsheet as the locus where such symmetry algebras emerge from the underlying closed string description, thereby sharpening the connection between causal structure, observer dependence, and string dynamics. Along this direction, following \cite{Bagchi:2024qsb}, it is worthwhile to revisit this framework from the perspective of an open string worldsheet from the outset. Furthermore, since our construction provides a first step toward the quantum description of finite-lifetime tensionless strings, it is important to investigate how it connects to existing approaches to quantizing closed and eternal null strings and to the emergence of three null strings \cite{Bagchi:2020fpr}. More broadly, the finite-lifetime perspective suggests a new paradigm in which tensionless strings are intrinsically tied to horizon-like structures and observer-accessible regions, paralleling the behavior of strings near spacetime horizons \cite{Bagchi:2021ban,Bagchi:2023cfp}. In this broader sense, the present framework elevates the tensionless limit from a purely asymptotic construction to an intrinsic physical phase of the worldsheet. This, in turn, opens avenues for exploring entanglement structure, near-horizon physics, connections to thermal phenomena, and the role of finite-lifetime induced boundaries in defining null string-theoretic phases.

\begin{acknowledgments}
SK acknowledges support from the IIT Guwahati postdoctoral fellowship (Project No.\ IITG/R\&D/IPDF/2023-24/25-0083).
\end{acknowledgments}






\makeatletter
\makeatother
\bibliographystyle{JAPS}
\bibliography{myref}

\end{document}